\documentclass[english,prd,superscriptaddress,nofootinbib,preprintnumbers,showpacs,preprint]{revtex4}
\usepackage[latin1]{inputenc}
\usepackage{graphicx}
\usepackage{amsmath,amsthm,amssymb}
\usepackage{bm}

\usepackage{amsfonts}
\usepackage{dcolumn}

\usepackage{color}

\renewcommand{\r}{\widetilde{r}}

\newcommand{\beqa}{\begin{eqnarray}}
\newcommand{\eeqa}{\end{eqnarray}}

\newcommand{\siml}{\lesssim}

\usepackage{babel}
\begin{document}

\title{A Note on Geodesics in Hayward Metric
}

\author{Takeshi Chiba}
\affiliation{Department of Physics, College of Humanities and Sciences, 
Nihon University, Tokyo 156-8550, Japan}
\author{Masashi Kimura}
\affiliation{ CENTRA, Departamento de Fisica, Instituto 
Superior Tecnico-IST, Universidade de Lisboa-UL, Avenida Rovisco Pais 1, 1049 Lisboa, Portugal}

\begin{abstract}
We study timelike and null geodesics in a non-singular black hole metric proposed by Hayward. 
The metric contains an additional length-scale parameter $\ell$ and approaches 
the Schwarzschild  metric at large radii while  
approaches a constant at small radii so that the singularity is resolved. 
We tabulate the various critical values of $\ell$ for timelike and null geodesics: the critical values for 
the existence of horizon,  marginally stable circular orbit and photon sphere. 
We find the photon sphere exists  even if the horizon is absent and 
two marginally stable circular orbits appear if the photon sphere is  absent 
 and a stable circular orbit for photons exists for a certain range of $\ell$.  
We visualize the image of a black hole  and find that blight rings appear even if the photon 
sphere is absent. 
\end{abstract}

\date{\today}


\maketitle

\section{Introduction}

A non-singular model of a black hole was proposed by Hayward \cite{hayward}. 
As such, it may be a solution of an ultraviolet complete gravity such as \cite{biswas}. 
The geodesics in such a non-singular spacetime would illuminate some 
basics aspects of the spacetime, as 
 the geodesics in the Schwarzschild spacetime do \cite{chandra}. 

In this paper, we summarize the results of our study of the properties of geodesics 
in the geometry described by 
the Hayward metric in a self-contained manner. 
Some of the results may be already known (e.g., \cite{wei,schee}), 
but the detailed study of both the properties of marginally stable circular orbits and 
the behaviour of null geodesics  are 
 new as far as we know. 
 
 In Sec. \ref{sec2}, after giving  the radii of the horizons of the spacetime, 
we study the properties of timelike 
 geodesics and null geodesics. Sec. \ref{sec3} is devoted to summary. 
 In Appendix \ref{appendixa}, we summarize the properties of the geodesics in the Reissner-Nordstr\"om 
 metric. 

We use the units of $G=c=1$. 

\section{Geodesics in Hayward Metric}
\label{sec2}

The Hayward metric is given by \cite{hayward}
\beqa
ds^2&=&-F(r)dt^2+\frac{dr^2}{F(r)}+r^2d\Omega^2\\
&&F(r)=1-\frac{2Mr^2}{r^3+2\ell^2M} \nonumber \,,
\label{metric}
\eeqa
where $M$ is a mass parameter and $\ell$ is a length-scale parameter. 
The metric approaches $1-2M/r$ as $r\rightarrow \infty$ and 
approaches unity smoothly as $r\rightarrow 0$ and hence is non-singular. 
The metric is ``minimal'' in the sense that it contains the least number of 
free parameters ($\ell$ only)
with the desired properties (regularity at the center such that $F(r)\rightarrow 1+{\cal O}(r^2)$ 
and Schwarzschild asymptotic behaviour at 
large radii. In fact, Frolov has recently shown that if $F(r)$ is a rational function of $r$, 
the order of the polynomials must be larger than  2 and an 
$F(r)$ constructed out of 
 polynomials of  order 3 contains two free parameters in general \cite{frolov}. 

We may compute the effective energy momentum for the metric Eq. (\ref{metric}) via 
$T_{\mu\nu}=\frac{1}{8\pi}G_{\mu\nu}$ as done in \cite{hayward}. \footnote{
The method for computing $T_{\mu\nu}$ in this way is 
sometimes called  ``Nariai method" in Japanese GR community.} 
Then, the energy density $\rho=-{T^t}_t$, the radial pressure $p_r={T^r}_r$ and 
the tangential pressure $p_T={T^{\theta}}_{\theta}={T^{\phi}}_{\phi}$ are given by \cite{hayward}
\beqa
\rho=-p_r&=&\frac{3\ell^2M^2}{2\pi(r^3+2\ell^2M)^2},\\
p_T&=&\frac{3\ell^2M^2(r^3-\ell^2M)}{\pi(r^3+2\ell^2M)^3}\, ,
\eeqa
and we find that the weak energy condition is satisfied: $\rho>0, \rho+p_r=0, \rho+p_T\geq 0$ 
and that  
the strong energy condition is violated for $r< (\ell^2M)^{1/3}$ since $\rho+p_r+2p_T=2p_T$. 

Henceforth, we normalize the length scale by $M$ and introduce the  dimensionless parameter
\beqa
a\equiv \frac{\ell}{M}\,,
\eeqa
and use  $\r=r/M$ as a dimensionless variable. 

\subsection{Horizon}

\begin{figure}
\includegraphics[height=3.5in]{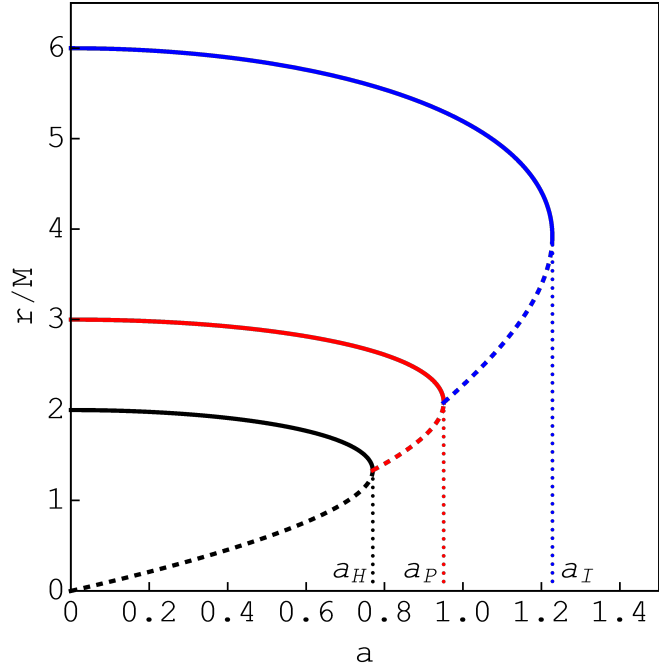}
\caption{\label{fig1} The radii of  horizons (black) (outer:  solid; inner:  dashed), 
ISCO or marginally stable circular orbit (blue), photon sphere (red  solid) and stable circular orbit of photon  (red  dashed). 
Dotted vertical lines are critical values of $a=\ell/M (a_H,a_P,a_I$ from left).  
 }
\end{figure}

The location of a horizon is determined by $F(r)=0$, and 
 horizons (outer horizon $\r_+$ and inner horizon $\r_-$) exist for  
$0\leq  a\leq a_{H}$ \cite{hayward}, 
\beqa
&&\r_+=\frac23 +\frac43 \cos\left(\frac13\cos^{-1}\left(1-\frac{27a^2}{8}\right)\right),\\
&&\r_-=\frac23 -\frac43 \cos\left(\frac13\cos^{-1}\left(1-\frac{27a^2}{8}\right)+\frac{\pi}{3}\right), 
\eeqa
where 
\beqa
a_H=\frac{4}{3\sqrt{3}}=0.7698...\,.
\eeqa
The upper bound on $a$ implies the lower bound on $M$, $M\geq a_H\ell$. The implications of 
this lower bound on the formation and the evaporation of black holes are discussed in \cite{hayward}. 
In Fig. \ref{fig1}, we show the radii of the horizons (black curve).
The causal structure is quite similar to the Reissner-Nordstr\"om black 
hole: $a<a_H,a=a_H, a>a_H$ corresponds to $Q<M, Q=M, Q>M$ 
Reissner-Nordstr\"om black hole, the only exception being  that 
the central singularity is replaced with a regular center.

\subsection{Timelike Geodesics}

{}From the spherical symmetry, we may restrict ourselves to the equatorial plane without loss of generality. 
In terms of two conserved quantities, the energy $E=F(r)\dot t$ and the angular momentum 
$L=r^2\dot\phi$, the timelike geodesics satisfy the  energy equation
\beqa
\frac12 \dot r^2+V(r)=\frac12 E^2,~~~~~~~~~~V(r)=\frac12\left(1+\frac{L^2}{r^2}\right)F(r)\,,
\eeqa
where $\dot t=dt/d\tau$ with $\tau$ being the proper time.

%

\begin{figure}
\includegraphics[height=2.5in]{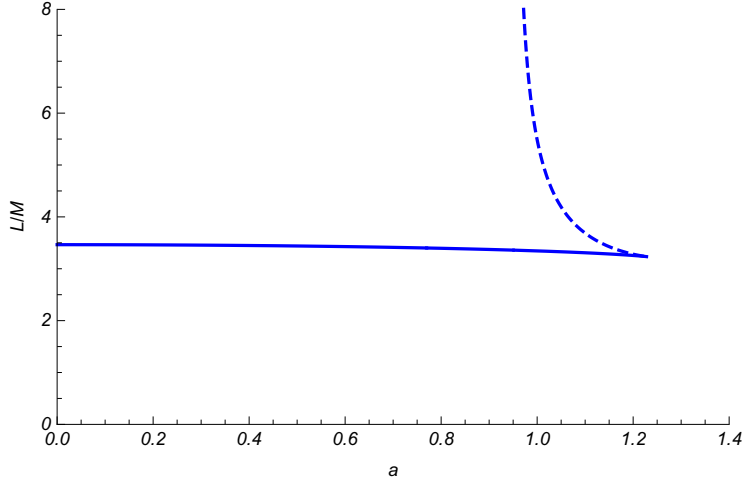}
\caption{\label{fig3} The angular momentum of marginally stable circular orbits. 
Solid (dashed) curve corresponds 
to blue solid (dashed) curve  in Fig. \ref{fig1}.  }
\end{figure}

A marginally stable circular orbit (MSCO) is determined by 
the condition $V'=V''=0$ which reduces to the following equation for $r$ 
\footnote{If $F(r)=1-Q_{n-1}(r)/P_n(r)$, where $P_n(r)$ and $Q_{n-1}(r)$ are polynomials of order $n$ and $n-1$, respectively, the degree of the equation becomes $3n-3$.}
\beqa
\r^6-6\r^5+22 a^2 \r^3-32a^4=0 \,,
\label{isco}
\eeqa
and the innermost such an orbit is called  the innermost stable circular orbit (ISCO).  
Then, $L^2$ is determined by 
\beqa
\left(\frac{L}{M}\right)^2=\frac{\r^3{dF}/{d\r}}{2F-\r {dF}/{d\r}}=\frac{\r^7-4 a^2 \r^4}{4 a^4+4 a^2 \r^3+(\r-3) \r^5}\,.
\eeqa 
We find that Eq. (\ref{isco}) has one positive real solution for $a< a_H$ and 
three positive real solutions for $a_H<a<a_I$ and one positive real 
solution for $a>a_I$, where  $a_I$ is given by  
\beqa
a_{I}=\frac{200}{51}\sqrt{\frac{5}{51}}=1.2278... \,.
\eeqa
However, both two of the solutions for $a_H<a<a_P$ 
and one  solution for $a>a_P$ are turned out to 
have imaginary $L$ and are unphysical, where 
\beqa
a_P=\frac{25}{24}\sqrt{\frac{5}{6}}=0.9509...\,,
\label{ap}
\eeqa
 so there is no ISCO for $a>a_I$. 
After all, there is one ISCO for $a< a_P$  and two MSCOs for 
$a_P<a<a_I$ and no ISCO for $a>a_I$. 
In Fig. \ref{fig1} and Fig.\ref{fig3}, we show the radii of ISCO or MSCO (blue curve) and the angular momentum. 
The angular momentum 
of the inner MSCO for $a_H<a<a_I$ is arbitrary large as $a\rightarrow a_P$. 

\begin{figure}
\includegraphics[height=3in]{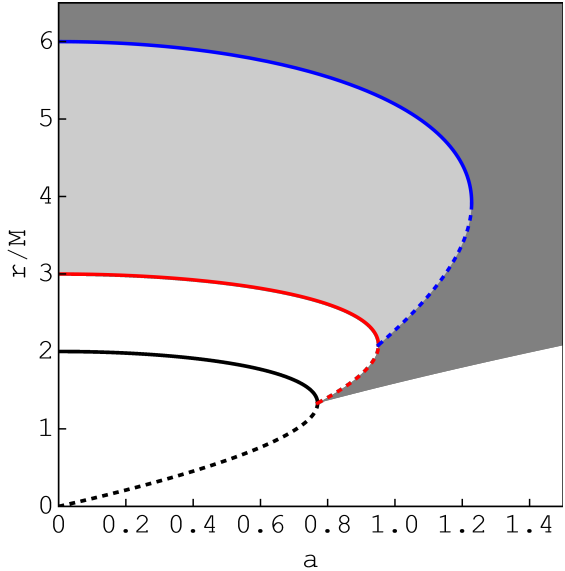}
\caption{\label{figcircle} The region of stable circular orbit (gray) as a function of $a$. Light gray
 region in the middle is the unstable  orbit. Lower white region 
is forbidden because either the angular momentum is imaginary or the radius is 
inside the horizon. Blue (ISCO), red (photon sphere 
and circular null geodesics) and black (horizons) curves are the same as Fig. \ref{fig1}. }
\end{figure}

In Fig. \ref{figcircle}, we show the allowed region for the stable circular orbit (gray). 
The light gray region in the middle is the unstable orbit $d^2V/dr^2<0$. 
The lower white  region is forbidden because the angular momentum is imaginary or  
the radius is inside the horizon.

\begin{figure}
\includegraphics[width=2.10in]{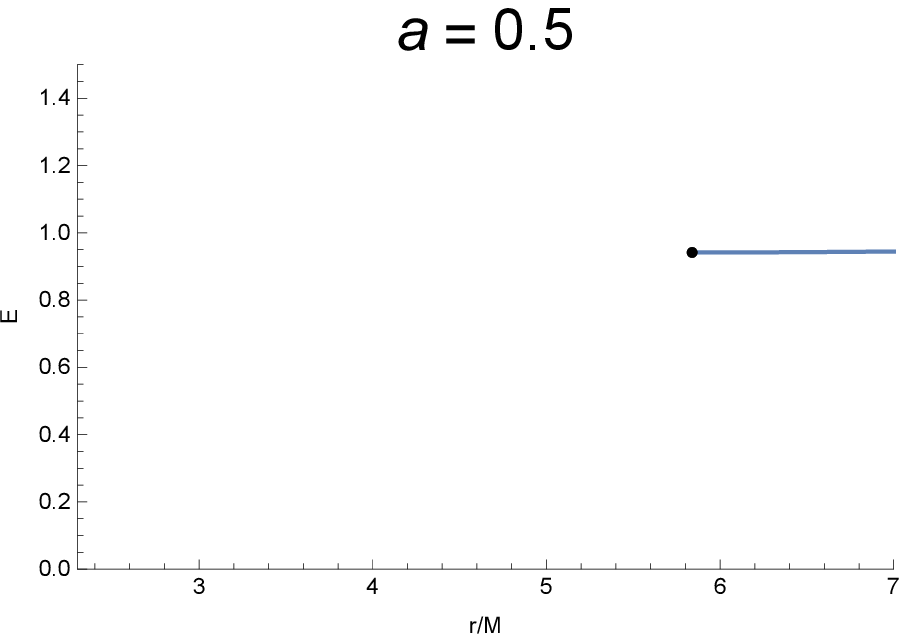}
\includegraphics[width=2.1in]{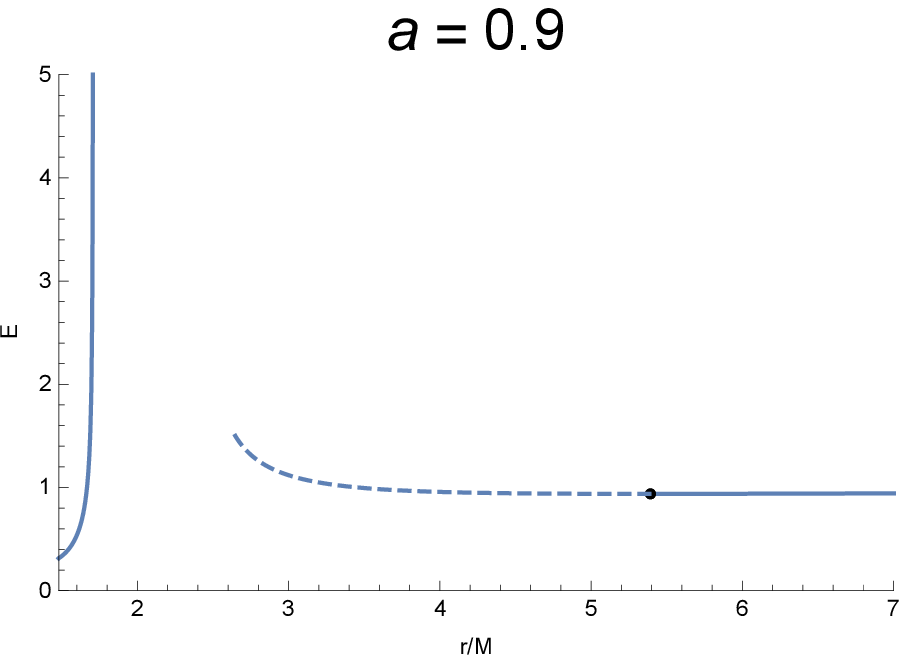}
\includegraphics[width=2.1in]{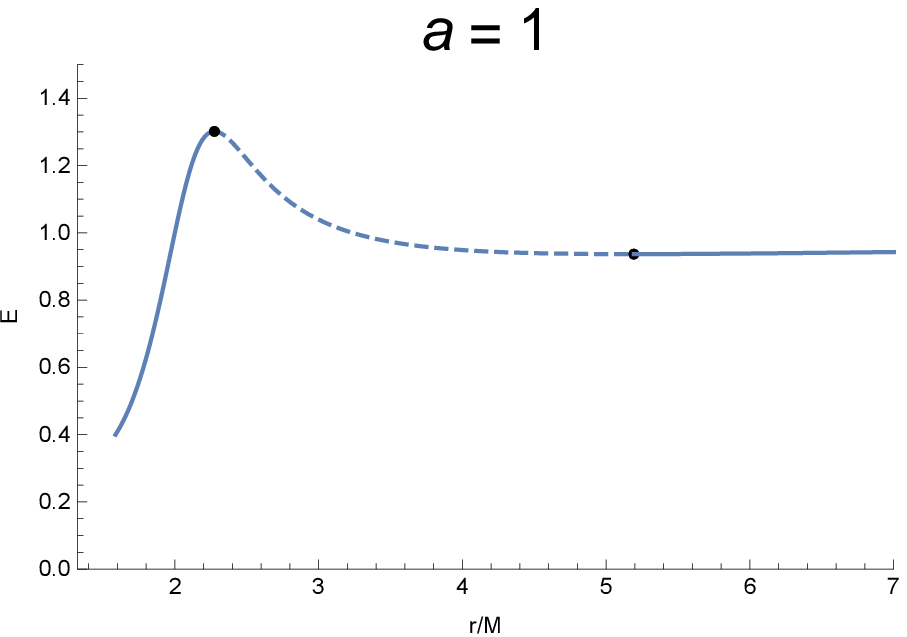}
\caption{\label{fig31} The energy  of a particle in circular orbit as a function of the radius 
of the orbit for $a=0.5,0.9,1$ from left to right. 
Black dots  correspond to marginally stable circular orbits. 
The circular orbit is unstable for dashed curve.  
  }
\end{figure}

In order to study the properties of circular orbits,  we calculate 
the energy of a particle in circular orbit as a function 
the radius for $a=0.5,0.9,1$ as shown in Fig. \ref{fig31} .  
The extrema of the energy correspond to MSCOs: 
for $a=1$ the minimum is the outer MSCO and the maximum is the inner MSCO. 
The region in between two MSCOs  
(dashed curve) is unstable because $d^2V/dr^2<0$ there.  
The outer MSCO can be reached from a particle 
in circular orbit outside by its emitting energy and angular momentum. 
On the other hand,  since the energy of the inner MSCO is larger than $1$ 
(the energy at large $r$),  the inner MSCO is unbound and may be regarded as an ``excited state" that 
cannot  reached from a particle 
in circular orbit outside.  In this sense, only the outer MSCO may be physically important. 
For $a=0.9$, there is also a stable circular orbit with large $L$ and  $E>1$.

\subsection{Null Geodesics}

As in the case of the timelike geodesics, in terms of $E$ and $L$, the null geodesics satisfy
\beqa
\frac12 \dot r^2+V_{\rm null}(r)=\frac12 E^2,~~~~~~~~V_{\rm null}(r)=\frac{L^2}{2r^2}F(r)\,, 
\label{energy:null}
\eeqa
where $\dot r=dr/d\lambda$ with $\lambda$ being the affine parameter. 

\begin{figure}
\includegraphics[height=2.2in]{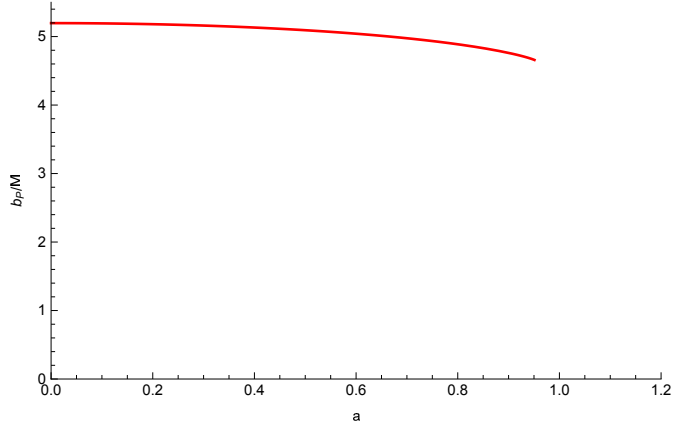}
\caption{\label{fig4} Shadow radius $b_{P}$ as a function of $a=\ell/M$.  
 }
\end{figure}

Since $L/E$ is the impact parameter at large $r$, 
the (local)  maximum of the effective potential $V_{\rm null}(r)$ determines the capture 
cross section for photons and hence the size of shadow of a black hole: photons with 
smaller impact parameter  will be captured by the black hole. 
The location of the (local) maximum of $V_{\rm null}(r)$, or the radius of unstable circular 
orbits of photons (photon sphere \cite{photon}), $r_P$, is determined by 
\beqa
 \r^6-3\r^5+4a^2\r^3+4a^4=0\,.
\eeqa
Similarly to the case of ISCO, we find that  the maximum of $V_{\rm null}(r)$  
exists if $a < a_P$, where $a_P$ is defined in Eq. (\ref{ap}). 
In Fig. \ref{fig1} we show the radius of photon sphere (red solid).  
There also appears a  stable circular orbit of 
photon  (red  dashed) for $a<a_P$. The appearance of a stable circular photon orbit is 
inevitable in the presence of the photon sphere (local maximum of $V_{\rm null})$) since 
$V_{\rm null}$ positively diverges as $r\rightarrow 0$. 
\footnote{The presence of a stable photon orbit, 
however,  would be problematic because perturbations can become long-lived and nonlinear effects may destabilize the system \cite{cardoso}. }
In Fig. \ref{fig4}, we show the radius of the shadow $b_{P}$ defined by 
\beqa
b_{P}=\sqrt{\frac{L^2}{2V_{\rm null}(r_{P})}}=\frac{r_P}{\sqrt{F(r_P)}}
\eeqa

In order to study the effect of the geometry on the propagation of light rays, 
 we first compute the deflection of light in the Hayward metric. 
We consider the deflection of light  in the equatorial plane. Then from 
Eq. (\ref{energy:null}) and $L=r^2\dot\phi$, 
in terms of the impact parameter $b=L/E$,  up to the turning point of the deflection, 
$\phi$ satisfies
\beqa
r^2\frac{d\phi}{dr}=\left(b^{-2}-\frac{F(r)}{r^2}\right)^{-1/2}\,.
\eeqa
Denoting the turning point of the deflection by $r_0$ at which $V_{\rm null}(r_0)=E^2/2$ 
or $b^{-2}=F(r_0)/r_0^2$, 
the deflection angle $\Delta\phi$ 
is then given by
\beqa
\Delta\phi=2\int^{1/r_0}_0\frac{du}{\sqrt{b^{-2}-u^2F(1/u)}}-\pi ,
\eeqa
where we have made the change of variables $u=1/r$ as usual. 
Expanding in terms of $M/r_0$ under the assumption $\Delta\phi$ is small (weak deflection), 
the result is
\beqa
\Delta\phi&=&\frac{4   M}{{r_0}}+\frac{(15 \pi
   -16) }{4}   \left(\frac{M}{r_0}\right)^2+
\frac{(244-45 \pi ) }{6} \left(\frac{M}{r_0}\right)^3+
\left(-130+\frac{3465\pi}{64} -\frac{15\pi}{4}a^2\right)
\left(\frac{M}{r_0}\right)^4\nonumber\\
&&+\left(\frac{7783}{10}-\frac{3465\pi}{16} +\frac{75 \pi
   -472  }{5}a^2\right)
    \left(\frac{M}{r_0}\right)^5\nonumber\\
&&+
\left(\frac{310695 \pi   }{256}-\frac{21397}{6} +\frac{5}{16} (1664-693   \pi )  a   ^2\right) \left(\frac{M}{r_0}\right)^6
+O\left(\left(\frac{M}{r_0}\right)^7\right) .
\eeqa
We show the result only up to $O((M/r_0)^6)$, although we can calculate arbitrarily higher orders. 
The coefficients for $a=0$ fully agree with \cite{keeton}, 
but the sign  of the coefficient of $(M/r_0)^4$ 
including $a$ is different from   
 \cite{wei} in which only the numerical values of the coefficient are given and  
  higher-order terms are not given.  
In terms of the impact parameter $b$ using $b^{-2}=F(r_0)/r_0^2$, 
the deflection angle up to $O((M/b)^6)$ is given by
\beqa
\Delta\phi&=&\frac{4 M}{b}+\frac{15 \pi  }{4}\left(\frac{M}{b}\right)^2
+\frac{128 }{3 }\left(\frac{M}{b}\right)^3
+\left(\frac{3465
   \pi  }{64}-\frac{15\pi}{4}    a^2\right)\left(\frac{M}{b}\right)^4
\nonumber\\
&&+\left(\frac{3584 }{5}-\frac{512
    a ^2}{5}\right)\left(\frac{M}{b}\right)^5
+\left(\frac{255255 \pi 
   }{256}-\frac{3465\pi}{16}   a^2\right)\left(\frac{M}{b}\right)^6
+
O\left(\left(\frac{M}{b}\right)^7\right) .
\eeqa
Again, we find  complete agreement with \cite{keeton} up to this order if $a=0$. 
Since the effect of nonzero $a$ appears only for  
$O((M/b)^4)$ and beyond,  it is difficult to detect the effect by the (weak) deflection angle.  
{}From the requirement that the fourth term should not dominate the first term 
by $0.001\%$ \cite{nature}, $a$ is constrained only as $a\siml 10^{5}(b/R_{\odot})^{3/2}(M/M_{\odot})^{-3/2}$ or $\ell\siml 10^{5}{\rm km}(b/R_{\odot})^{3/2}(M/M_{\odot})^{-1/2}$.

\begin{figure}
\includegraphics[width=1.55in]{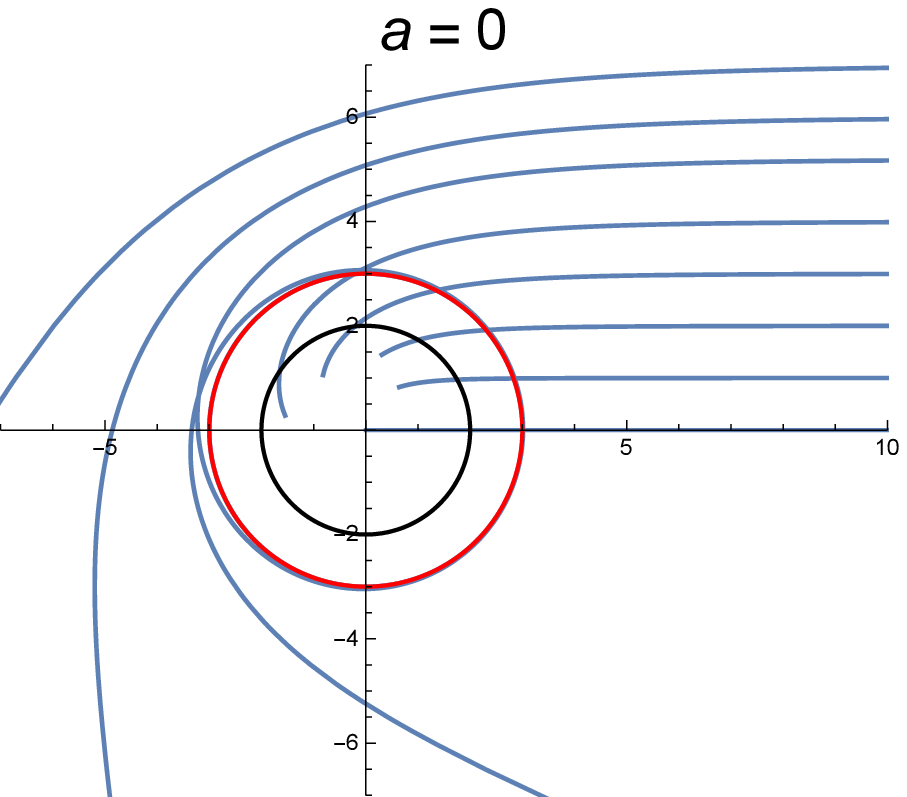}
\includegraphics[width=1.55in]{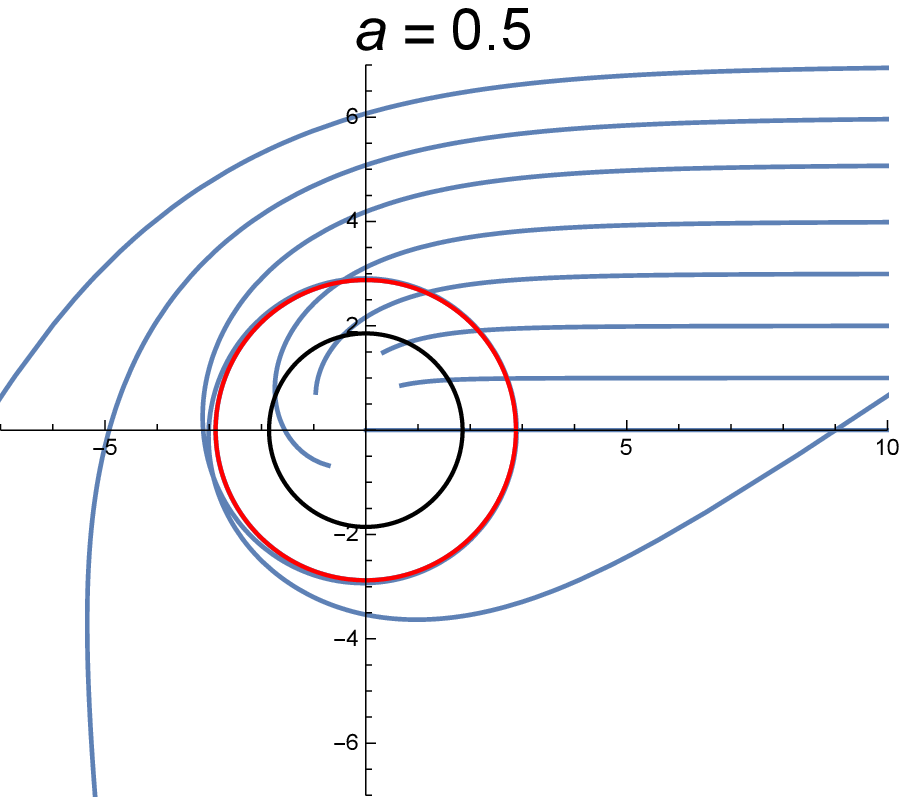}
\includegraphics[width=1.55in]{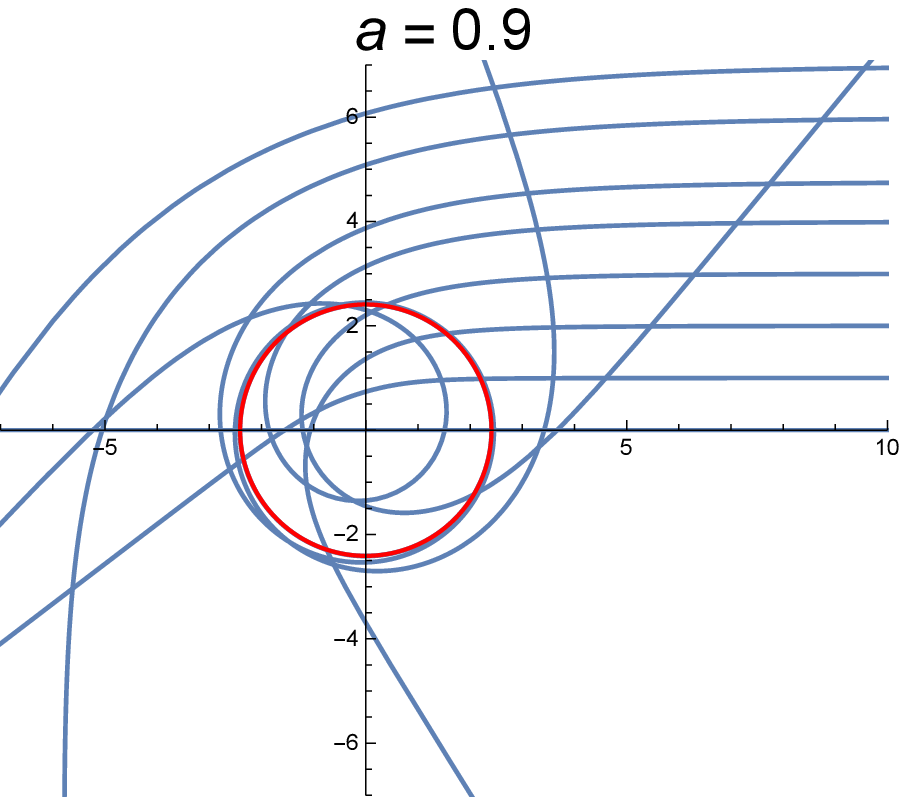}
\includegraphics[width=1.55in]{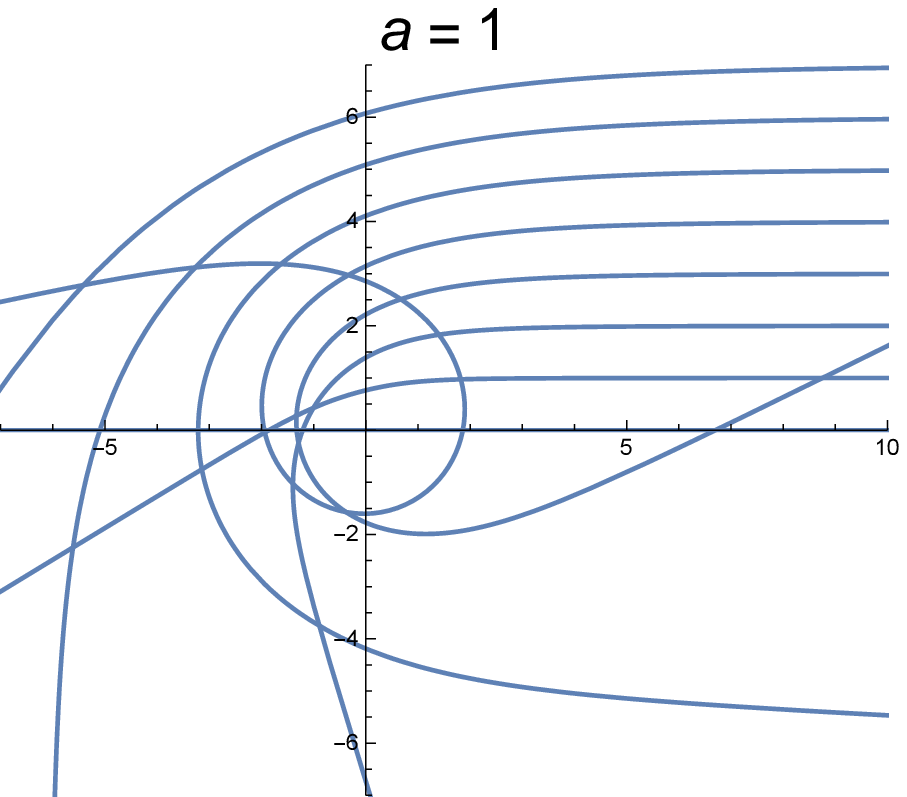}
\caption{\label{fig5} Trajectories of light rays (coming in from upper right) 
for $a=0, 0.5, 0.9, 1$ from left to right.  A black circle is the 
(outer) horizon, red circles are the radii of photon spheres.  
 }
\end{figure}
\begin{figure}
\includegraphics[width=1.55in]{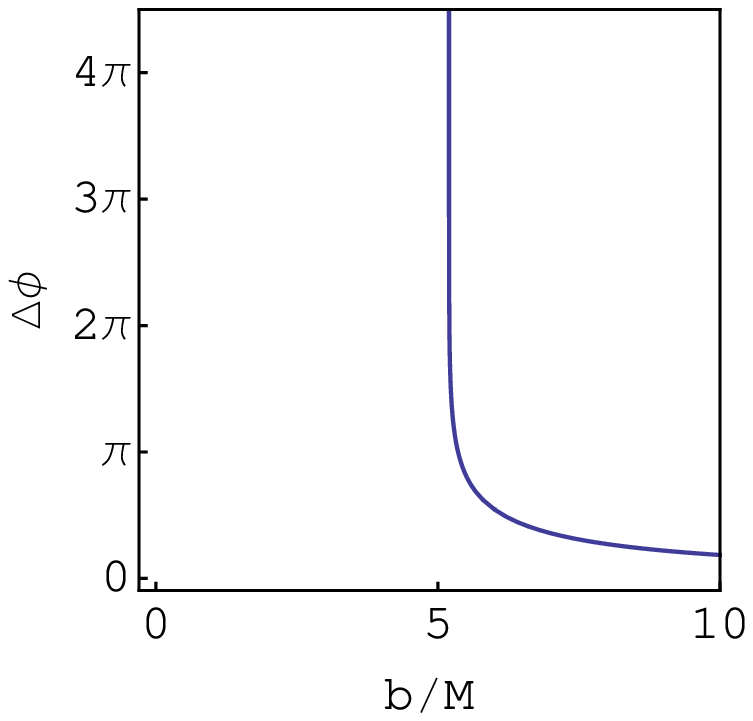}
\includegraphics[width=1.55in]{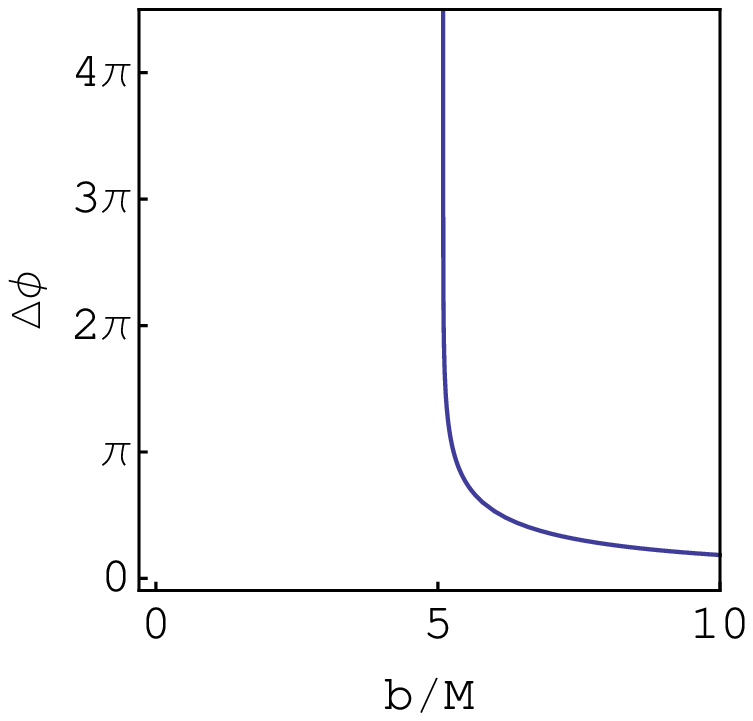}
\includegraphics[width=1.55in]{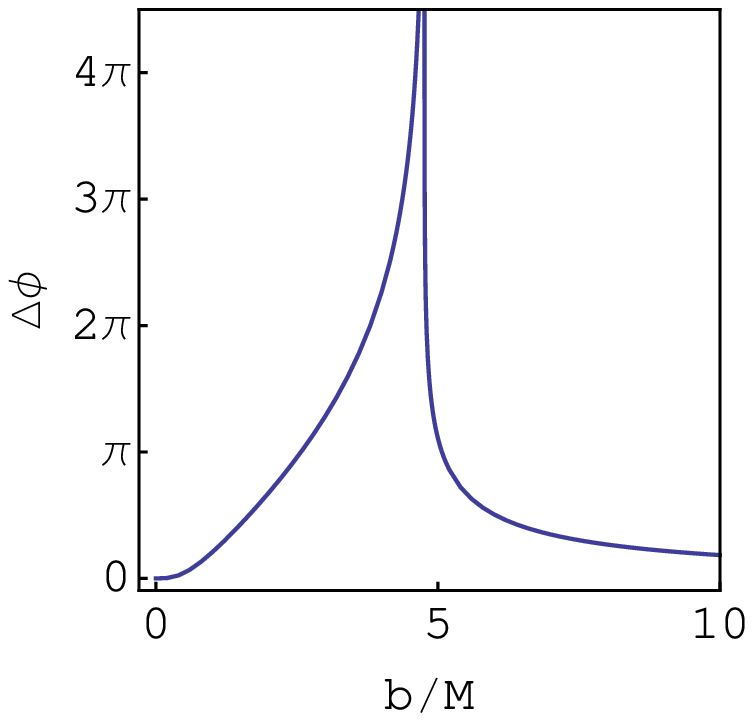}
\includegraphics[width=1.55in]{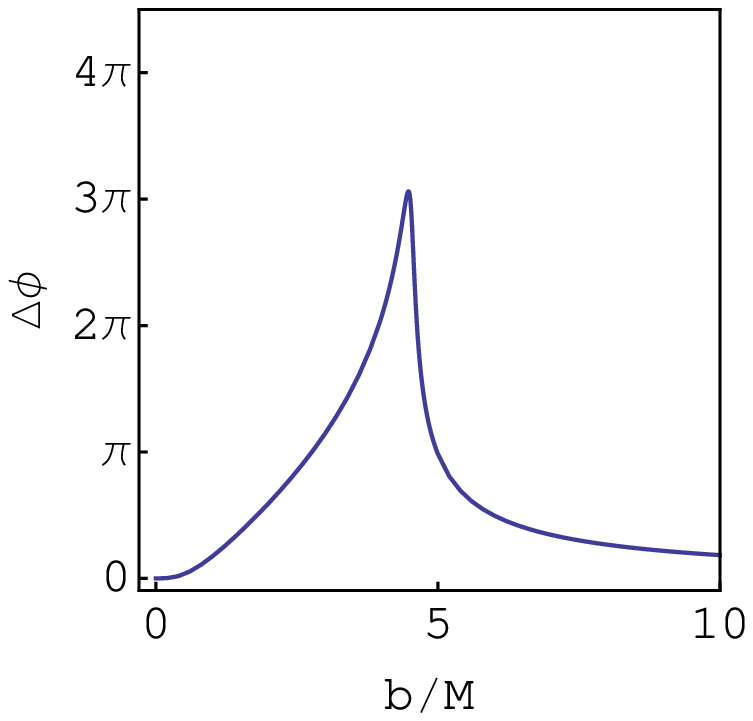}
\caption{\label{delphi} 
The deflection angle as a function of the impact parameter $b$ 
for $a=0, 0.5, 0.9, 1$ from left to right.
 }
\end{figure}
\begin{figure}
\includegraphics[height=1.55in]{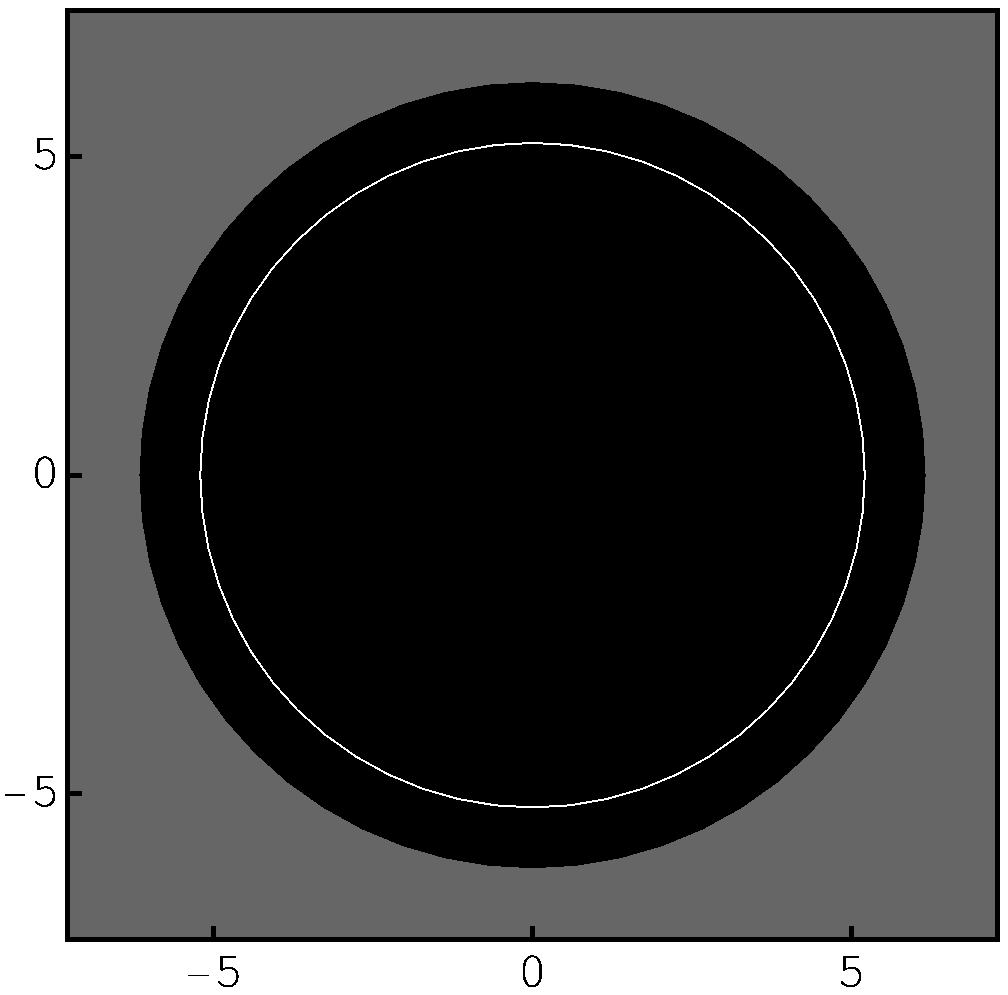}
\includegraphics[height=1.55in]{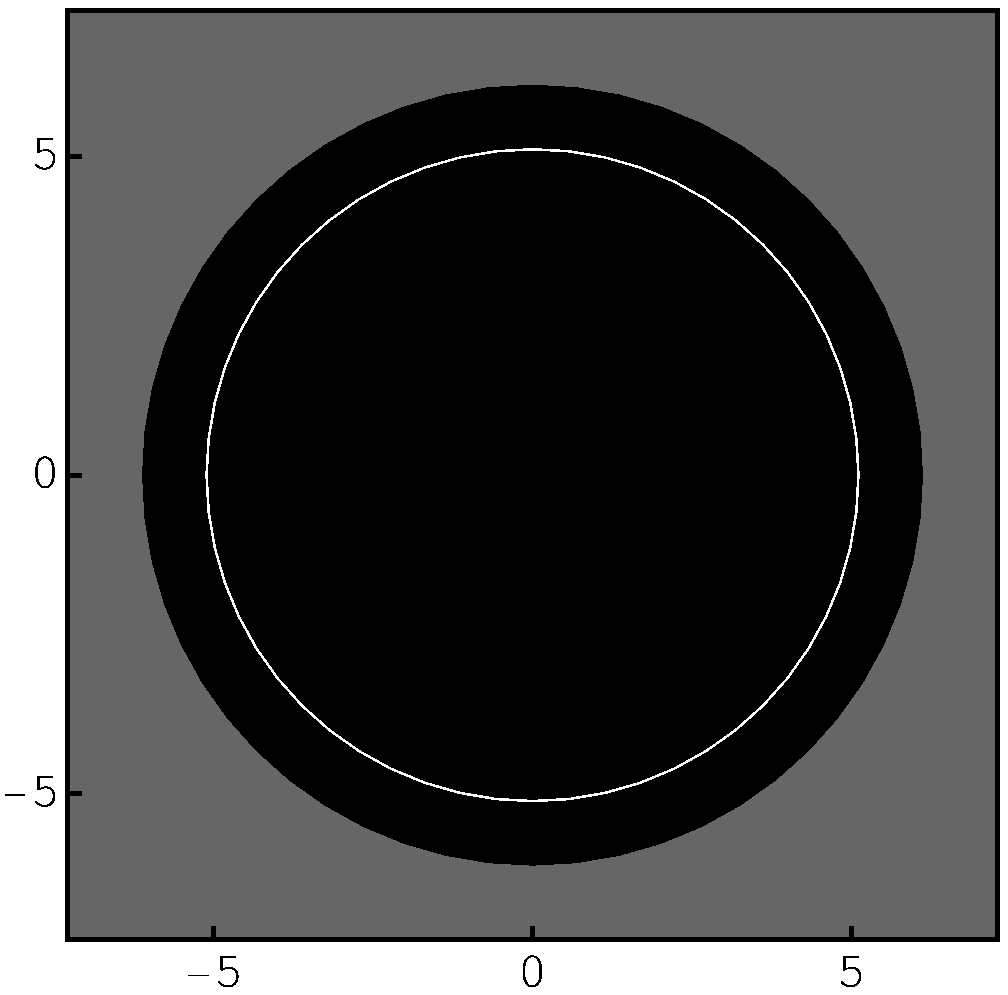}
\includegraphics[height=1.55in]{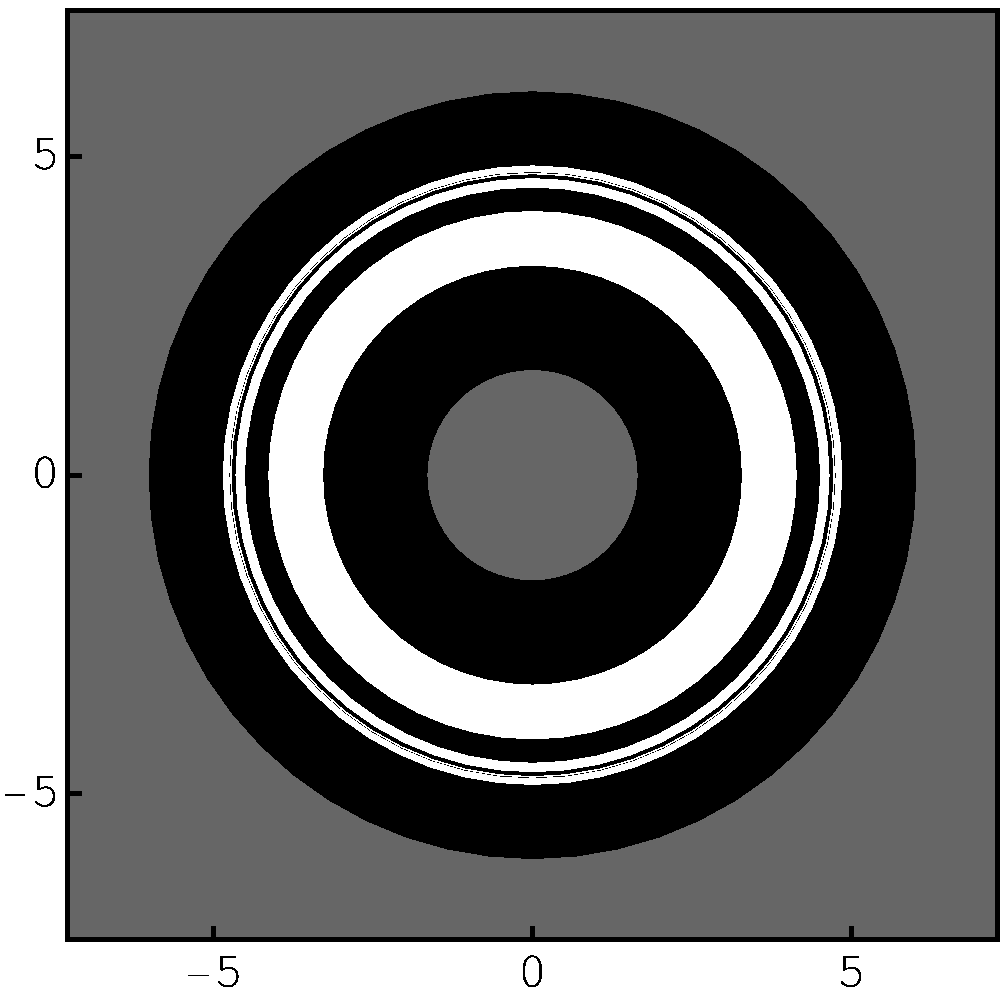}
\includegraphics[height=1.55in]{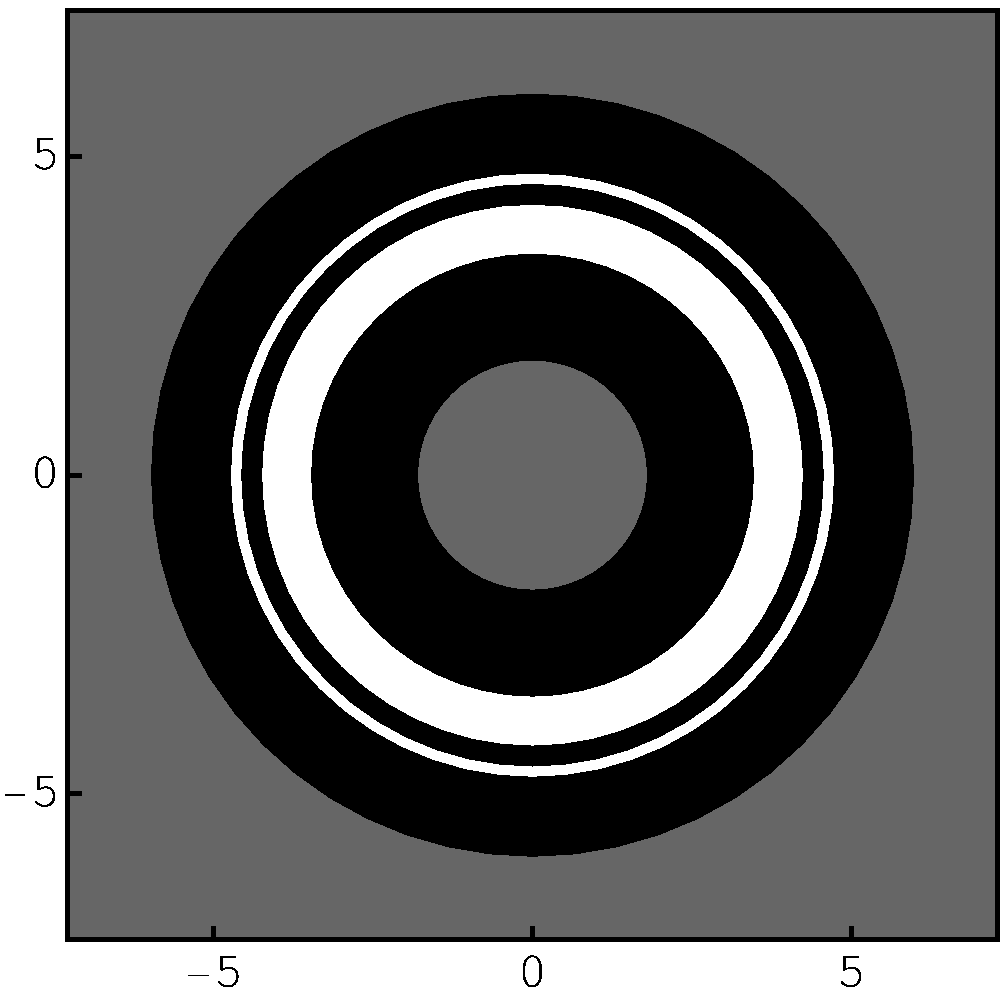}
\caption{\label{shadow} Image of a non-singular black hole described by the Hayward metric for $a=0, 0.5, 0.9, 1$ from left to right. 
  Photons in black regions never reach the observer. Photons in white regions 
  travel around the central region more than once, while photons in  
gray regions reach the observer straightforwardly. 
 }
\end{figure}

However, the presence of $a$ does affect the existence or nonexistence of the photon sphere and hence 
affects the behaviour of light rays traveling around the photon sphere. 
In Fig. \ref{fig5}, we  show the trajectories of light rays coming 
in from upper right for $a=0,0.5,0.9,1$. A black circle is the 
(outer) horizon, red circles are the radii of photon spheres. 
In Fig. \ref{delphi}, we show the deflection angle as a function of the impact parameter $b$  
for $a=0,0.5,0.9,1$. 
If the impact parameter is slightly larger than the radius of a shadow $b_{P}$, 
the deflection angle $\Delta\phi$ is logarithmically divergent as 
\beqa
\Delta\phi & \sim &  \frac{1}{\sqrt{-5 + 2\sqrt{3 r_P/M}}} 
\ln(b - b_{P})^{-1},
\eeqa
If $a = a_P$, {\it i.e.,} $r_P = 25/12$, the divergent behavior is changed into 
\beqa
\Delta\phi & \sim & c_0 M^{1/6}(b - b_{P})^{-1/6},
\\
c_0 & = & 2^{11/6} 3^{2/3} 5^{5/4} \int_0^\infty dy\frac{1}{\sqrt{432y + 900y^2 +625 y^3}} \simeq 7.771.
\eeqa
Quite similar behaviour is also found for the Reissner-Nordstr\"om metric 
where critical values corresponding to $a_H,a_P$ appear depending on the value of the charge 
(Appendix\ref{appendixa}). 

The images of a ``black hole" are shown in Fig \ref{shadow}. 
Photons in black regions 
never reach the observer due to the presence of a horizon or the deflection angle 
being $\pm \pi/2$(modulo $2\pi$). Photons in  white regions travel around the central region more than once, while photons in   gray regions reach the observer straightforwardly. 
The image for $a=0.5$ is little different from that of Schwarzschild black hole ($a=0$): 
A blight ring surrounding a black disk (shadow of a black hole) appears. 
Interestingly,  for $a=0.9 (a_H<a<a_P)$, the photon sphere exists 
even though  the horizon is absent;  a black disk 
disappears and a black doughnut appears instead, and a ring image persists. 
More interestingly, even for $a=1 (>a_P)$, the ring persists although the photon  
sphere is absent as shown in Fig. \ref{shadow}. 
Therefore, the existence of a blight ring image does not necessarily imply 
the existence of a photon sphere.  
Of course, for astrophysical black holes $a\simeq 10^{-38}(\ell/\ell_{\rm p})(M_{\odot}/M)\ll 1$, 
 where $\ell_{\rm p}$ is the Planck length,   
 and these phenomena would be expected only for Planck-scale
primordial black holes.

\section{Summary}
\label{sec3}

In this paper, we have studied the timelike and null geodesics in a non-singular black hole 
geometry proposed by Hayward that involves a parameter $\ell$  
and found several interesting features of the geometry: 
the existence or non-existence of the horizon,  the 
photon sphere and the ISCO. 
We also have found that two marginally stable circular orbits appear for $a_H<\ell/M<a_P$ 
although the inner orbit is unbound. 
The existence  of a horizon and/or a photon sphere  
significantly affects the behaviour of the null geodesics. 
We have found that a black doughnut appears if the horizon is absent and that 
blight rings appear even if the photon sphere is absent. 

One may think that  bright regions in a shadow image 
are due to the existence of the photon sphere. 
However, as shown in Fig.\ref{shadow}, bright regions 
also can appear in the spacetime with no photon sphere 
if the parameter $a$ is slightly larger than $a_P$.
For such parameters, while there are no photon sphere, $\Delta \phi$ can still take 
a value larger than $2\pi$ (see Fig.\ref{delphi}).
Similar behavior can be found in the Reissner-Nordstr\"om metric.
These results suggest that such the shadow images are universal behavior for parameters
slightly larger than the critical value where photon sphere marginally exists.
The parameter $\ell$ is currently only loosely constrained  by the solar-system experiment 
since the deflection angle at the post-Newtonian order is the same as  
that of Schwarzschild.  The Event Horizon Telescope, a long baseline interferometer experiment, will be able to resolve black holes at horizon scales 
with the angular resolution of $20 \mu as$ which corresponds to a size of 
$3.6 M$ for SgA* at the galactic center \cite{eht}. 
If the shadow of a black hole is observed by such a telescope, 
we may put an  $O(1)$ constraint on  $\ell/M$.

\section*{ACKNOWLEDGEMENTS}

This work is supported by the MEXT Grant-in-Aid for Scientific Research on
 Innovative Areas (15H05894) 
and in part
by Nihon University. 
M.K. acknowledges financial support provided under the European Union's H2020 ERC 
Consolidator Grant ``Matter and strong-field gravity: New frontiers in Einstein's theory''
grant agreement no. MaGRaTh-646597, and under the H2020-MSCA-RISE-2015 Grant No. StronGrHEP-690904.

\appendix


\section{Reissner-Nordstr\"om metric}
\label{appendixa}

The Reissner-Nordstr\"om metric is given by
\beqa
ds^2&=&-F(r)dt^2+\frac{dr^2}{F(r)}+r^2d\Omega^2\\
&&F(r)=1-\frac{2M}{r} + \frac{Q^2}{r^2} \nonumber \,,
\label{RNmetric}
\eeqa
where $M, Q$ are mass and electric charge parameters, respectively.
We study the properties of geodesics on this metric 
and compare the result with the case of the Hayward metric.

Since the electric charge parameter appears in the metric as $Q^2$, 
we need to consider $Q\ge 0$ only.
If $1 \ge Q/M$, there are two horizons at $r = r_{\pm} = M^2 \pm \sqrt{M^2 - Q^2}$.
\begin{figure}
\includegraphics[height=2.7in]{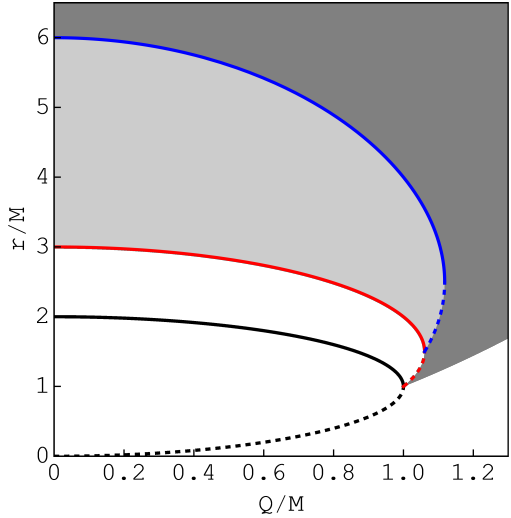}
\caption{\label{figcircleRN} 
The region of stable and unstable circular orbits as a function of $Q/M$.
}
\end{figure}
In Fig.\ref{figcircleRN}, we have plotted the regions of stable and 
unstable circular orbits. 
The qualitative feature is same as the case of the Hayward metric Fig.\ref{figcircle}.
When $Q/M$ takes $Q_H/M = 1, 
Q_P/M = 3/(2\sqrt{2}) \simeq 1.061, 
Q_I/M = \sqrt{5}/2 \simeq 1.118$, 
the horizon becomes extremal, the photon sphere marginally exists, the ISCO marginally exists, respectively.
The locations of the photon sphere and the shadow image for 
the Reissner-Nordstr\"om metric are also discussed in~\cite{Zakharov:2014lqa}.

\begin{figure}
\includegraphics[width=1.55in]{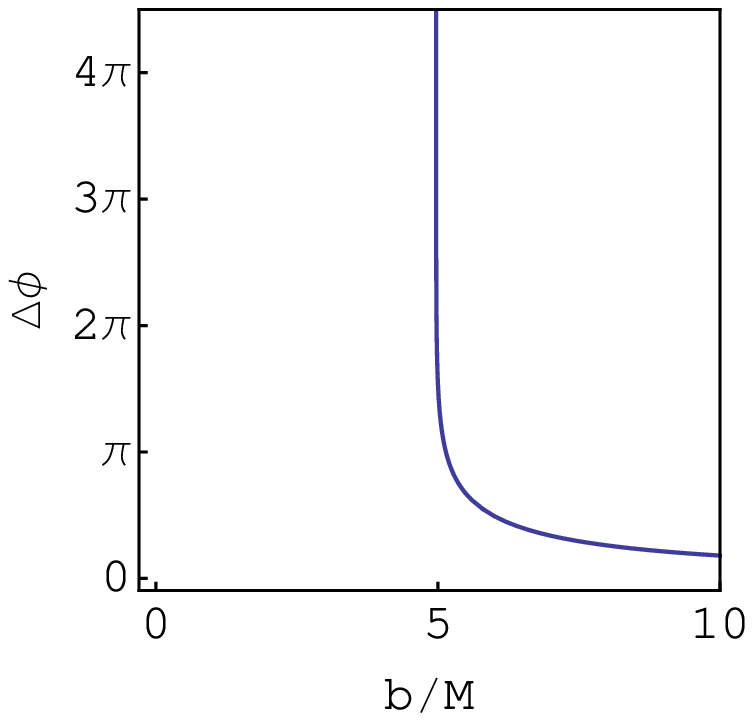}
\includegraphics[width=1.55in]{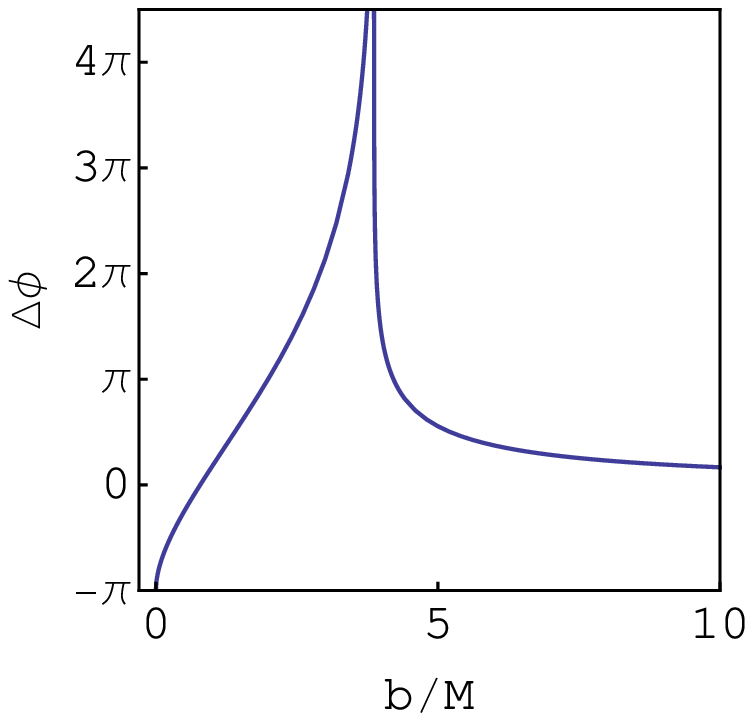}
\includegraphics[width=1.55in]{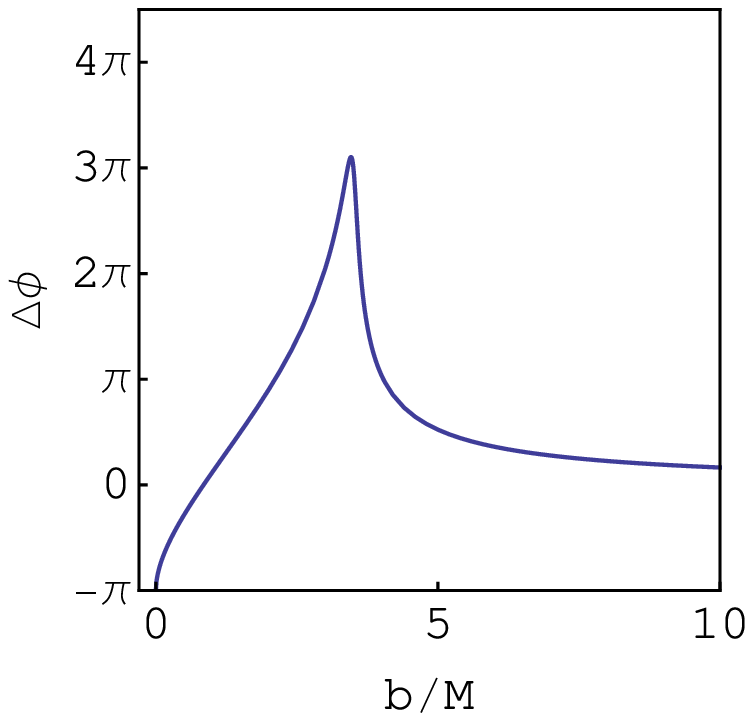}
\caption{\label{delphiRN} 
The deflection angle as a function of the impact parameter $b$ 
for $Q/M = 0.5, 1.03, 1.08$ from left to right.
 }
\end{figure}
Fig.\ref{delphiRN} shows the deflection angle for null geodesics
as a function of the impact parameter $b$.
Compared to the case of the Hayward metric (see Fig.\ref{delphi}), 
the qualitative behavior is almost the same.
The only difference can be found near the region for small impact parameter 
$b \simeq 0$ in overcharged cases $Q/M > 1$, 
where the deflection angle $\Delta \phi$ is almost $-\pi$.
This is because the null geodesic is reflected
by the repulsive force near the origin $r = 0$.

If we consider $Q > Q_P$, there is no photon sphere. 
However, for a parameter value slightly larger than $Q_P$ (see the third figure in Fig.\ref{delphiRN}),
$\Delta \phi$ can still take a value larger than $2\pi$.
This means that their shadow image is similar to the fourth figure in Fig.\ref{shadow}.



\begin{thebibliography}{10}

\bibitem{hayward}
S.~A.~Hayward,
  Phys.\ Rev.\ Lett.\  {\bf 96}, 031103 (2006)
  doi:10.1103/PhysRevLett.96.031103
  [gr-qc/0506126].

\bibitem{biswas}
L.~Modesto,
  Phys.\ Rev.\ D {\bf 86}, 044005 (2012)
  [arXiv:1107.2403 [hep-th]]; 
T.~Biswas, E.~Gerwick, T.~Koivisto and A.~Mazumdar,
  Phys.\ Rev.\ Lett.\  {\bf 108}, 031101 (2012)
  doi:10.1103/PhysRevLett.108.031101
  [arXiv:1110.5249 [gr-qc]].
  
\bibitem{chandra}
S. Chandrasekhar, {\it The Mathematical Theory of Black Holes}, 
{\it Oxford University Press (1983)}.


\bibitem{wei}
S.~W.~Wei, Y.~X.~Liu and C.~E.~Fu,
  Adv.\ High Energy Phys.\  {\bf 2015}, 454217 (2015)
  doi:10.1155/2015/454217
  [arXiv:1510.02560 [gr-qc]].
  
\bibitem{schee}
J.~Schee and Z.~Stuchlik,
  JCAP {\bf 1506}, 048 (2015)
  doi:10.1088/1475-7516/2015/06/048
  [arXiv:1501.00835 [astro-ph.HE]].
  

\bibitem{frolov} 
V.~P.~Frolov,
  Phys.\ Rev.\ D {\bf 94}, no. 10, 104056 (2016)
  doi:10.1103/PhysRevD.94.104056
  [arXiv:1609.01758 [gr-qc]].

\bibitem{photon}
K.~S.~Virbhadra and G.~F.~R.~Ellis,
  Phys.\ Rev.\ D {\bf 62}, 084003 (2000)
  doi:10.1103/PhysRevD.62.084003
  [astro-ph/9904193].


\bibitem{cardoso}
V.~Cardoso, L.~C.~B.~Crispino, C.~F.~B.~Macedo, H.~Okawa and P.~Pani,
  Phys.\ Rev.\ D {\bf 90}, no. 4, 044069 (2014)
  doi:10.1103/PhysRevD.90.044069
  [arXiv:1406.5510 [gr-qc]].
  
\bibitem{keeton}
C.~R.~Keeton and A.~O.~Petters,
  Phys.\ Rev.\ D {\bf 72}, 104006 (2005)
  doi:10.1103/PhysRevD.72.104006
  [gr-qc/0511019].

\bibitem{nature}
B.~Bertotti, L.~Iess and P.~Tortora,
  Nature {\bf 425}, 374 (2003).

\bibitem{eht}
S.~Doeleman {\it et al.},
  Nature {\bf 455}, 78 (2008)
  doi:10.1038/nature07245
  [arXiv:0809.2442 [astro-ph]].


\bibitem{Zakharov:2014lqa} 
  A.~F.~Zakharov,
  Phys.\ Rev.\ D {\bf 90}, no. 6, 062007 (2014)
  doi:10.1103/PhysRevD.90.062007
  [arXiv:1407.7457 [gr-qc]].

\end{thebibliography}
\end{document}